\documentclass[a4paper,oneside]{amsart}
\usepackage{latexsym}
\usepackage{amsmath}
\usepackage{amssymb}
\let\amslrcorner\lrcorner 
\usepackage{amsfonts}
\usepackage{mathrsfs}
\usepackage{mathabx}
\usepackage{tensor}
\usepackage{float}
\usepackage{esint}
\usepackage{enumitem}
\usepackage{graphicx}
\usepackage{color}
\usepackage[pagebackref=false,bookmarks=false,pdfborder={0 0 0}]{hyperref}
\let\lrcorner\amslrcorner  

\allowdisplaybreaks 
\definecolor{Blue}{rgb}{0.,0.,1.}
\definecolor{Red}{rgb}{1.,0.,0.}
\definecolor{Green}{rgb}{0.,1.,0.}
\let\origmaketitle\maketitle
\def\maketitle{
  \begingroup
  \def\uppercasenonmath##1{} 
  \let\MakeUppercase\relax 
	\origmaketitle
  \endgroup
	}

\newcounter{smallarabics}
\newenvironment{arabicenumerate}
{\begin{list}{{\normalfont\textrm{(\arabic{smallarabics})}}}
  {\usecounter{smallarabics}\setlength{\itemindent}{0cm}
   \setlength{\leftmargin}{5ex}\setlength{\labelwidth}{4ex}
   \setlength{\topsep}{0.75\parsep}\setlength{\partopsep}{0ex}
   \setlength{\itemsep}{0ex}}}
{\end{list}}

\newcounter{smallroman}

\newcommand{\ben}{\begin{arabicenumerate}}  
\newcommand{\een}{\end{arabicenumerate}}

\def\init{\setcounter{equation}{0}}

\newtheorem*{thmchap}{Theorem}
\newtheorem{theoreme}{Theorem}[section]
\newtheorem{proposition}[theoreme]{Proposition}
\newtheorem{hypothesis}[theoreme]{Hypothesis}
\newtheorem{lemma}[theoreme]{Lemma}
\newtheorem{definition}[theoreme]{Definition}
\newtheorem{corollary}[theoreme]{Corollary}
\theoremstyle{definition} 
\newtheorem{remark}[theoreme]{Remark}
\newtheorem{example}[theoreme]{Example}
\newcommand{\beq}{\begin{equation}}
\newcommand{\eeq}{\end{equation}}

\newcommand{\bex}{\begin{example}}
\newcommand{\eex}{\end{example}}
\def\bel{\begin{lemma}}
\def\eel{\end{lemma}}
\def\bet{\begin{theoreme}}
\def\eet{\end{theoreme}}
\def\bed{\begin{definition}}
\def\eed{\end{definition}}
\def\ber{\begin{remark}}
\def\eer{\end{remark}}

\def\rr{{\mathbb R}}

\def\cc{{\mathbb C}}
\def\nn{{\mathbb N}}

\def\ss{{\mathbb S}}

\def\diff{d}

\def\part{{\rm par}}

\let\Im\relax
\let\Re\relax
\DeclareMathOperator{\Im}{Im}
\DeclareMathOperator{\Re}{Re}

\def\bar{\overline}

\def\cinf{C^\infty}
\def\c0inf{C_0^\infty}

\def\proof{
\noindent{\bf Proof.}\ \ }

\def\cD{{\mathcal D}}

\def\cB{{\mathcal B}}

\def\cf{C^\infty}

\def\i{{\rm i}}

\def\qed{$\Box$\medskip}

\def \p{ \partial}

\def\12{\frac{1}{2}}
\def\14{\frac{1}{4}}

\newcommand{\one}{\boldsymbol{1}}

\def\coinf{C_{\rm c}^\infty}

\def \p{ \partial}

\def\12{\frac{1}{2}}

\newcommand{\mat}[4]{\left(\begin{array}{cc}#1 &#2  \\ #3 &#4 \end{array}\right)}

\newcommand{\col}[2]{\left(\begin{array}{c}#1 \\#2\end{array} \right)}

\def\CARal{{\rm C\hskip 0.25 em \hbox{\raise 1.72 ex 
\hbox{$\scriptscriptstyle\rm al$}\kern -0.57 em A}R}}

\def\otimesal{\mathop{\hbox{\raise 1.5 ex
  \hbox{$\scriptscriptstyle\rm al$}
\kern -0.92 em \hbox{$\otimes$}}}}
\def\oplusal{\mathop{\hbox{\raise 1.5 ex
  \hbox{$\scriptscriptstyle\rm al$}
\kern -0.92 em \hbox{$\oplus$}}}}
\def\Gammal{\hbox{\raise 1.68 ex 
\hbox{$\scriptscriptstyle\rm al$}\kern -0.50 em $\Gamma$}}
\def\Bal{\hbox{\raise 1.68 ex 
\hbox{$\scriptscriptstyle\rm  al$}\kern -0.50 em $B$}}
\def\CARal{{\rm C\hskip 0.25 em \hbox{\raise 1.72 ex 
\hbox{$\scriptscriptstyle\rm al$}\kern -0.57 em A}R}}

\newcommand{\traa}[1]{\mskip-6mu\upharpoonright_{#1}}
\newcommand{\tra}[1]{\mskip-6mu\upharpoonright_{#1}}

\def\WFA{{\rm WF}_{\rm a}}
\makeatletter
\newcommand*{\defeq}{\mathrel{\rlap{%
                     \raisebox{0.3ex}{$\m@th\cdot$}}%
                     \raisebox{-0.3ex}{$\m@th\cdot$}}%
                     =}
\makeatother
\makeatletter
\newcommand*{\eqdef}{=\mathrel{\rlap{%
                     \raisebox{0.3ex}{$\m@th\cdot$}}%
                     \raisebox{-0.3ex}{$\m@th\cdot$}}%
                     }
\makeatother

\newcommand{\bea}{\begin{aligned}}
\newcommand{\beal}{\begin{array}{l}}
\newcommand{\eeal}{\end{array}}
\newcommand{\eea}{\end{aligned}}
\newcommand{\bec}{\begin{cases}}
\newcommand{\eec}{\end{cases}}

\def\cN{\mathcal{N}}

\DeclareMathOperator{\Ran}{Ran}

\DeclareMathOperator{\supp}{supp}

\def\mo{\mathscr{O}}

\def\calde{Calder\'{o}n }
\def\wavefront{wave front }

\def\zero{{\rm\textit{o}}}

\def\b2{\frac{\beta}{2}}

\def\rk{{\bf k}}\def\rh{{\bf h}}\def\rg{{\bf g}}
\def\cB{\mathcal{B}}
\def\cN{\mathcal{N}}
\def\Me{X}

\def\mo{\mathscr{O}}

\def\calde{Calder\'{o}n }
\newenvironment{notations}
{\begin{list}{{\normalfont\textrm{-}}}
  {\setlength{\itemindent}{0cm}
   \setlength{\leftmargin}{2ex}\setlength{\labelwidth}{4ex}
   \setlength{\topsep}{0.75\parsep}\setlength{\partopsep}{1ex}
   \setlength{\itemsep}{1ex}}
}
{\end{list}}

\newcommand{\dred}[1]{}

\def\pOmega{\p\Omega}
\def\ccf{C_{\rm c}^{\infty}}
\def\is{\kappa}
\def\cs{{\rm c}}
\def\gl{{\rm gl}}
\def\loc{{\rm loc}}
\DeclareMathAlphabet{\mathpzc}{OT1}{pzc}{m}{it}

\def\ovk{\overline{\rk}}

\newcommand{\opencl}[1]{\mathopen{}\mathclose{\left]#1 \right]}}

\newcommand{\open}[1]{\mathopen{}\mathclose{\left]#1 \right[}}

\begin{document}

\title[Wick rotation of the time variables for two-point functions on analytic backgrounds]{\LARGE Wick rotation of the time variables  \\ for two-point functions on analytic backgrounds}
\author{\large Micha{\l} Wrochna
\address{Universit\'e de Cergy-Pontoise, D\'epartement de Math\'ematiques, 2 avenue Adolphe Chauvin, 95302 Cergy-Pontoise \textsc{Cedex}, France}
\email{michal.wrochna@u-cergy.fr}}

\thanks{\emph{Acknowledgments.} The author is very grateful to Christian G\'erard for inspiring discussions and helpful suggestions, and would also like to thank the anonymous referees for their help in improving the paper. Support from the grant ANR-16-CE40-0012-01 is gratefully acknowledged.}

\begin{abstract} We set up a general framework for Calder\'{o}n projectors (and their generalization to non-compact manifolds), associated with complex Laplacians e.g.~obtained by Wick rotation of a Lorentzian metric. In the analytic case, we use this to show that the Laplacian's Green's functions have analytic continuations whose boundary values are two-point functions of analytic Hadamard states. The result does not require the metric to be stationary. As an aside, we describe how thermal states are obtained as a special case of this construction if the coefficients are time-independent.
\end{abstract}

\maketitle


\section{Introduction and summary} 

\subsection{Introduction} One of the cornerstones of Quantum Field Theory on Min\-ko\-wski space is the analytic continuation of Euclidean theories down to the Lorentzian signature known as the \emph{Wick rotation} {\cite{OS,GJ}}. 

Its generalization to \emph{stationary}  spacetimes (in particular \emph{static} spacetimes) is a highly natural and efficient tool which lies at the heart of many important results in the construction of distinguished propagators and states, see e.g.~\cite{wald,fulling,sewell,sanders1,sanders,HHI2,DS},  as well as in renormalization problems, see e.g.~\cite{wald,moretti1,HM}.  

On the other hand, outside of the stationary case there is no obvious, canonical procedure generalizing the Wick rotation that is directly useful to all of these applications simultaneously. This makes the study of Quantum Field Theory on non-stationary backgrounds particularly difficult, much in contrast to the vast array of techniques available in the Euclidean case, see e.g.~\cite{BD,dang,DDR} for recent advances. Despite the lack of a universal proposal, techniques based on various approaches to Wick rotation have been successfully applied to several important problems on non-stationary spacetimes. For instance, a local procedure was used by Moretti in the context of Hadamard-parametrix based renormalization \cite{moretti}, and a closely related analysis of the Hadamard parametrix was employed by Sanders to prove the Hadamard property in his construction of the Hartle-Hawking-Israel state \cite{sanders}.  A different procedure based on a Wick-rotated elliptic problem was proposed earlier by Candelas and Raine \cite{CR} for the construction of Feynman parametrices, and a more detailed analysis has been recently used for  global index computations on perturbations of Minkowski space by Gell-Redman, Haber and Vasy \cite{GHV}.

Recently, a new proposal has been made in the context of constructing linear Klein-Gordon quantum fields via their \emph{two-point functions}. Building on an idea due to G\'erard \cite{HHI}, it was shown in \cite{analytic} that pairs of two-point functions $\Lambda^\pm\in\cD'(M^2)$ on an analytic spacetime $(M,\rg)$ can be constructed in terms of \emph{\calde projectors} for a Wick-rotated elliptic problem. Under global hyperbolicity assumptions for $(M,\rg)$, it turned out possible to control the properties required of two-point functions, i.e.~their \emph{positivity} and the microlocal \emph{Hadamard condition}. Furthermore, this provided the first general proof of existence of two-point functions satisfying an analytic version of the Hadamard condition, proposed originally by Strohmaier, Verch and Wollenberg \cite{SVW}, and shown by them to imply the so-called \emph{Reeh-Schlieder property} (for applications see the recent review \cite{witten} on entanglement in QFT).

The construction in \cite{analytic} made use of analytically continuing the Klein-Gordon operator $P$ to its elliptic counterpart, denoted here by $K$. However, the objects directly relevant for QFT, namely the Lorentzian two-point functions $\Lambda^\pm$ and the elliptic inverse $K^{-1}$, were connected  from the outset only in an indirect way, through Cauchy data of the former and Calder\'on projectors associated with the latter. 

The main goal of the present work is to show that $\Lambda^\pm$ and the Schwartz kernel of $K^{-1}$ (the \emph{Euclidean Green's function}) are actually directly related by Wick rotation, rather than merely in an `instantaneous' way.

We do not assume that the metric is stationary, so in this generality it is not possible to analytically continue in the difference of the two time variables. Our results are also valid in a more general setting than in \cite{analytic}, where Wick rotation was tied to the choice of Gaussian normal coordinates, and $K^{-1}$ was constructed by imposing Dirichlet boundary conditions close to the Wick-rotated Cauchy surface. Inspired by ideas from the recent work \cite{HHI2}, we use a more covariant framework for the \calde projectors, and we also allow for more general operators that are not projectors. Furthermore, rather than restricting ourselves to specific boundary conditions, we work with an abstract set of properties required of $K^{-1}$, which can be met in various different situations. Our main motivation was to include cases like \emph{thermal states} in the form in which they are interpreted in G\'erard's approach to the construction of the Hartle-Hawking-Israel state \cite{HHI,HHI2}, where periodic boundary conditions have to be imposed (cf.~the earlier work of Sanders \cite{sanders}).
 
\subsection{Main result} We consider a globally hyperbolic spacetime $(M,\rg)$, which is a neighborhood of $\{t=0\}$ in $\rr_t\times \Sigma$, with $\Sigma$ a real analytic manifold, and where $\rg$ is a real-analytic metric of the form
\[
\rg= - N^{2}(t,y) dt^{2}+ \rh_{jk}(t,y)(dy^{j}+ w^{j}(t,y)dt)(dy^{k}+w^{k}(t,y)dt).
\]
We assume that in the first variable, it analytically continues {to a complex domain}, and on the imaginary axis it defines a \emph{complex metric}
\[
\rk =N^{2}(\i s,y)ds^{2}+ \rh_{jk}(\i s,y)(dy^{j}+ \i w^{j}(\i s,y)d s)(dy^{k}+ \i w^{k}(\i s,y)d s).
\]
Assuming $|\rk(x)| = {\rm det}(\rk_{ab}(x))>0$ we consider the Klein-Gordon operator and its Wick-rotated counterpart:
\[
P= -|\rg|^{-\12}\p_{a}|\rg|^{\12}\rg^{ab}\p_{b}+\mu, \quad  K=  - |\rk|^{-\12}\p_{a}\rk^{ab}|\rk|^{\12}\p_{b}+\lambda,
\]
where {$(t,y)\mapsto\mu(t,y)$ is real-analytic} and $\lambda(s,y)=\mu(\i s,y){=\overline{\mu(-\i s,y)}}$. Note that $K$ is not necessarily formally self-adjoint; instead, its formal adjoint $K^*$ w.r.t.~$|\rk|^{\12}ds dy$ equals $\kappa\circ K \circ \kappa$, where $(\kappa u)(s,y)=u(-s,y)$. We need $K$ to be \emph{elliptic} on a $\kappa$-invariant neighborhood $X$ of $\{s=0\}$ in $\rr_s\times\Sigma$ and \emph{invertible} on a suitable domain, which is taken care of by assuming Hypothesis \ref{hyp:coercive} and Hypothesis \ref{hyp:K} (with $\Omega^\pm=\{\pm s> 0\}\cap X$) in the main part of the text. With these assumptions, disregarding the spatial variables when writing Schwartz kernels,  we have:

\begin{thmchap}[{cf.~Thm.~\ref{thm:main} for more detailed formulation}] 
There exists $\delta>0$, a pair of two-point functions $\Lambda^\pm$ for $P$, and a pair of holomorphic functions $F^\pm$ of two variables with values in $\cD'(\Sigma^2)$ such that:
\beq\label{eq:dfgsdf}
\bea
K^{-1}(s_1,s_2)&= F^\pm(\i s_1, \i s_2), \ \ \pm s_1>0, \ \pm s_2<0,\\
\Lambda^\pm (t_1,t_2)&= F^\pm\big((t_1,t_2)\pm\i \Gamma 0\big), \ \ t_1,t_2\in \open{-\delta,\delta},
\eea
\eeq
the last expression meaning the limit $(s_1,s_2)\to 0$ from $\pm\Gamma= \{\pm s_1>0, \ \pm s_2<0\}$.
\end{thmchap} 

Here, we adopt the terminology which is conventional for charged fields, see \cite{analytic} for a more detailed introduction. In particular, by \emph{two-points functions} we mean bi-solutions $\Lambda^\pm\in\cD'(M^2)$ of the Klein-Gordon equation which are Schwartz kernels of positive operators on $\ccf(M)$, and which satisfy $\Lambda^+-\Lambda^-=\i G$, with $G$ a prescribed bi-solution called the \emph{Pauli-Jordan} or \emph{causal propagator}. To fulfill those conditions,  we construct $\Lambda^\pm$ as in \cite{analytic} by specifying their Cauchy data at $t=0$ in terms of operators generalizing the \calde projectors associated with $K^{-1}$, and then the fact that $\Lambda^\pm$  are two-point functions boils down to a careful analysis of these operators. In particular, the property $\Lambda^\pm\geq 0$ is a consequence of a weak version of \emph{reflection positivity} at zero imaginary time. The proof of \eqref{eq:dfgsdf} makes use of the resulting formul\ae{} and their symmetries, combined with a generalized version of the method developed in \cite{analytic}, which in turn is a distributional version of an argument due to Schapira in the $D$-module setting \cite{Sch}.

\subsection{Summary} In Section \ref{sec:cald} we define and prove basic properties of what we call \emph{\calde operators} for complex Laplacians. The framework there is general and does not use analyticity. Section \ref{sec:wick} introduces the geometric setup for Wick rotation and describes how the two-point functions $\Lambda^\pm$ are obtained from \calde operators. Section \ref{sec:anal} contains the proof of the main result of the paper.  Finally, in Section \ref{sec:kms} we show that the KMS condition for $\Lambda^\pm$ arises as a simple corollary of periodic boundary conditions in the stationary case.


\section{\calde operators}\label{sec:cald}
\subsection{Complex metrics}\label{ss:complexmetrics} In the present section we will consider a geometric setup introduced recently by G\'erard \cite[Sec.~9.1]{HHI2}, and study operators that generalize the \emph{\calde projectors}, considered usually for real Laplacians on compact manifolds, see e.g.~\cite{Gr}. We also generalize the recent analysis from \cite{analytic,HHI2}.

If $X$ is a smooth manifold, we denote by ${\rm T}^{p}_{q}(X)$ the space of smooth, real valued $(p,q)$ {tensor fields} on $X$ and by $\cc {\rm T}^{p}_{q}(X)$ its complexification. A {\em complex metric} on $X$ is a symmetric and non-degenerate element $\rk= \rk_{ab}(x)dx^{a}dx^{b}$ of $\cc {\rm T}^{0}_{2}(X)$.

Let $\rk= \rk_{ab}(x)dx^{a}dx^{b}$ be a complex metric on an orientable smooth manifold $X$.  We will assume the following:
\begin{hypothesis}\label{hyp:coercive} For all $x\in X$, $|\rk(x)| = {\rm det}(\rk_{ab}(x))>0$, and {${v}_a \rk^{ab}(x) v_b \notin \opencl{-\infty,0}$} for all $v\in T_{x}^*X$, $v\neq 0$, {where $\rk^{cd}$ is the inverse matrix of $\rk_{ab}$}. 
\end{hypothesis}
\subsection{Laplacians}
We can define the Christoffel symbols:
\[
\Gamma^{c}_{ab}\defeq \12 \rk^{cd}(\p_{a}\rk_{bd}+ \p_{b}\rk_{ad}- \p_{d}\rk_{ab}),
\]
and the covariant derivative:
\[
\nabla^{(\rk)}_{a}T^{b}= \p_{a}T^{b}+ \Gamma_{ac}^{b}T^{c}.
\]
By a direct computation one can show that we have the same identity as in the pseudo-Riemannian case, namely:
\[
\nabla_{a}^{(\rk)}T^{a}= |\rk|^{-\12} \p_{a}(|\rk|^{\12}T^{a}),
\]
see \cite[Sec.~A.3]{HHI2}.

Let $\lambda\in \cinf(X)$. We consider the elliptic differential operator
\beq\label{eq:defK}
K\defeq - \nabla^{(\rk)}_{a}\rk^{ab}\nabla^{(\rk)}_{b}+\lambda= - |\rk|^{-\12}\p_{a}\rk^{ab}|\rk|^{\12}\p_{b}+\lambda.
\eeq
The differential operator $- \nabla^{(\rk)}_{a}\rk^{ab}\nabla^{(\rk)}_{b}$ is what one could call the \emph{Laplace-Beltrami operator associated to the complex metric $\rk$}. 

Note that the complex adjoint metric $\bar{\rk}$ satisfies $|\bar{\rk}|=|\rk|$. Denoting by $K^{*}$ the formal adjoint of $K$ for the scalar product
\[
(v|u)\defeq \int_{X}\bar{v}u|\rk|^{\12}dx,
\]
we obtain that:
\beq\label{eq:Kadjoint}
K^{*}= - |\overline{\rk}|^{-\12}\p_{a}\overline{\rk}^{ab}|\overline{\rk}|^{\12}\p_{b}+\bar{\lambda}= - \nabla^{(\ovk)}_{a}\ovk^{ab}\nabla^{(\ovk)}_{b}+\bar{\lambda}.
\eeq

\subsection{Gauss-Green formula} Let  $\Omega^+\subset X$ be an open set with a (not necessarily connected) smooth boundary. We denote the complement of its closure by
\[
\Omega^{-}\defeq X\backslash (\Omega^+)^{\rm cl}.
\]
We assume that the regions $\Omega^+$ and $\Omega^-$ have a common {smooth} boundary, which will be denoted by $\pOmega$.

In the complex setting, the \emph{outer unit normal vector field} to $\pOmega$, denoted by $n\in \cc TX$, is uniquely defined by the conditions:
\[
\begin{array}{rl}
a)& n(x)\cdot\rk(x)v= 0, \ \forall v\in T_{x}\pOmega, \\[2mm]
b)& n(x)\cdot \rk(x)n(x)=1,\\[2mm]
c)& \Re n(x) \hbox{ is outwards pointing}.
\end{array}
\]
Condition $a)$ singles out a single complex ray in $\cc T_{x}X$ and conditions $b)$ and $c)$ determine $n$ uniquely provided that $n(x)\cdot \rk(x)n(x)\neq 0$. To see that the latter condition is satisfied,  we can pick a boundary-defining function $f\in \cinf(X; \rr)$ for $\pOmega$, meaning that $\Omega^+$ is locally given by $\{f>0\}$ and $df\neq 0$ on $\{f=0\}$, and then construct $n$ using the formula
\[
n^{a}= \frac{-\rk^{ab}\nabla^{(\rk)}_{b}\!f}{\big((\nabla^{(\rk)}_{a}f)\rk^{ab}(\nabla^{(\rk)}_{b}\!f{)}\big)^{\12}}.
\]
In what follows we will also need the following condition:
\[\begin{array}{rl}
d)& \Im n(x)\in T_{x}\pOmega.
\end{array}
\]
This is equivalent to $(\nabla^{(\rk)}_{a}\!f)\rk^{ab}(\nabla^{(\rk)}_{b}\!f){\geq 0}$ on $\pOmega$. We will consider the following situation, in which {\it d)} holds: 

\begin{hypothesis}\label{hyp:kappa} Assume there exists a  diffeomorphism $\is: X \to X$
such that
\[
\is^{*}\rk= \ovk, \ \ \is^{*}\lambda=\overline\lambda, \ \ \is\circ \is= {\rm Id}, \ \ \is: \Omega^{{\pm}} \to \Omega^{{\mp}}
\]
homeomorphically, and $\kappa |_{\p\Omega}={\rm Id}$. 
\end{hypothesis}

We will denote by the same letter $\is$ the pullback $\is: \cD'(X)\to\cD'(X)$.

\begin{lemma} Hypothesis \ref{hyp:kappa} has the following consequences:
\beq\label{e1.8z}
\begin{array}{rl}
1)& D_{x}\is n(x) = - \bar{n}(x) \ \ \forall x\in \pOmega,\\[2mm]
2)& \Im n(x)\in T_{x}\pOmega \ \ \forall x\in \pOmega,\\[2mm]
3)& K^*=\kappa\circ K \circ \kappa. 
\end{array}
\eeq
\end{lemma}
\proof  We have $\bar{n}(x)\cdot\ovk(x) v=0$, $\bar{n}(x)\cdot\ovk(x)\bar{n}(x)=1$  and $\Re \bar{n}(x)$ is outwards pointing for $x\in \pOmega$. {To show 1), we prove $-D_{x}\is \bar{n}(x) = {n}(x)$ by checking that it satisfies properties $a)$--$c)$}.  Since $D_{x}\is= \one$ on $T_{x}\pOmega$ and $\is^{*}\rk= \ovk$, this implies that $D_{x}\is\bar{n}\cdot \rk(x)v=0$, $D_{x}\is\bar{n}\cdot \rk(x)D_{x}\is\bar{n}=1$. {Finally, since} $\is: \Omega^{+}\to \Omega^{-}$, $\Re D_{x}\is \bar{n}(x)$ is inwards pointing.

Next, we  have $v\in T_{x}\pOmega$ iff $D_{x}\is v=v$ and $D_{x}\is\Im n= \dfrac{1}{2\i}D_{x}\is(n - \bar{n})= \dfrac{1}{2\i}(-\bar{n}+ n)$. This proves 2). 

Finally, 3) is a direct corollary of $\is^{*}\rk= \ovk$, {$\is^{*}\lambda=\overline\lambda$} and \eqref{eq:Kadjoint}. \qed

We denote by $\rh$ the complex metric induced by $\rk$ on $\pOmega$. One can show that $\rh$  satisfies $|\rh|>1$. We denote by $d{\rm Vol}_{\rk}= |\rk|^{\12}dx^{1}\wedge \cdots \wedge dx^{n}$ the volume form on $X$, and by $d{\rm Vol}_{\rh}$ the volume form on $\pOmega$. The associated densities are $d\mu_{\rk}= | d{\rm Vol}_{\rk}|$ and $\diff\sigma_{\rh}= |d{\rm Vol}_{\rh}|$.

If $i: \pOmega\to X$ is the canonical injection, one can show that
\[
i^{*} (n\lrcorner\, d{\rm Vol}_{\rk})= d{\rm Vol}_{\rh}, \ \ i^{*}(n\lrcorner\,  d\mu_{\rk})= d \sigma_{\rh}.
\]
This allows one to derive the Gauss-Green formula:
 \beq\label{e1.0}
\int_{\Omega^+}\nabla_{a}^{(\rk)}V^{a}\diff\mu_{\rk}= \int_{\p \Omega}V^{a}n_{a}\diff\sigma_{\rh}.
\eeq

\subsection{Several identities} We have the following consequences of \eqref{e1.0}.

\begin{lemma} For any $u\in \cf(X)$ and $v\in\ccf(X)$, we have:
\begin{equation}
\label{e1.1z}
\int_{\Omega^+}(\bar{v}Ku- \overline{K^{*}v}u) \diff\mu_{\rk}= \int_{\p \Omega}(n^{a}\nabla_{a}^{(\rk)}\bar{v}u - \bar{v}n^{a}\nabla_{a}^{(\rk)}u)\diff\sigma_{\rh},
\end{equation}
\begin{equation}
\label{e1.4z}
\bea
\int_{\Omega^+}(\bar{v}Ku+ \bar{Kv}u)\diff\mu_{\rk}&=2\int_{\Omega^+}(\overline{\nabla_{a}^{(\rk)}v}\Re \rk^{ab}\nabla_{b}^{(\rk)}u +{(\Re\lambda)} \bar{v}u) \diff\mu_{\rk}\\[2mm]
&\phantom{=\,} -\int_{\p\Omega}( \overline{n^{a}\nabla^{(\rk)}_a v}u+ \bar{v}n^{a}\nabla_{a}^{(\rk)}u )\diff\sigma_{\rh}.
\eea\end{equation}
\end{lemma}
\proof For $V^{a}= \bar{v}\rk^{ab}\nabla^{(\rk)}_{b}u$ we obtain $\nabla^{(\rk)}_{a}V^{a}= \nabla^{(\rk)}_{a}\bar{v}\rk^{ab}\nabla^{(\rk)}_{b}u- \bar{v}Ku+\lambda\bar{v}u$, hence by \eqref{e1.0}:
\beq\label{e1.2z}
\int_{\Omega^+}\bar{v}Ku \diff\mu_{\rk}= \int_{\Omega^+}(\nabla^{(\rk)}_{a}\bar{v}\rk^{ab}\nabla^{(\rk)}_{b}u+\lambda\bar{v}u)\diff\mu_{\rk}- \int_{\p \Omega} \bar{v}n^{a}\nabla_{a}^{(\rk)}u\diff\sigma_{\rh}.
\eeq
The same identity with $\rk$ replaced by $\ovk$, $\lambda$ replaced by $\overline{\lambda}$ {and $u$ interchanged with $v$} gives:
\[
\int_{\Omega^+}\bar{u}K^{*}v \diff\mu_{\ovk}= \int_{\Omega^+}(\nabla^{(\ovk)}_{a}\bar{u}\ovk^{ab}\nabla^{(\ovk)}_{b}v+\bar\lambda \bar{u}v)\diff\mu_{\ovk}- \int_{\p \Omega} \bar{u}\bar{n}^{a}\nabla_{a}^{(\ovk)}v\diff\sigma_{\overline{\rh}},
\]
which by taking complex conjugates gives:
\beq\label{e1.3z}
\int_{\Omega^+}u\overline{K^{*}v} \diff\mu_{\rk}= \int_{\Omega^+}(\nabla^{(\rk)}_{a}u\rk^{ab}\nabla^{(\rk)}_{b}\bar{v}+\lambda u\bar{v})\diff\mu_{\rk}- \int_{\p \Omega} un^{a}\nabla_{a}^{(\rk)}\bar{v}\diff\sigma_{\rh},
\eeq
since $\overline{\nabla_{a}^{(\rk)}u}= \nabla_{a}^{(\ovk)}\bar{u}$. Subtracting \eqref{e1.3z} from \eqref{e1.2z} gives \eqref{e1.1z}. 

Let us now show \eqref{e1.4z}. Since $d\mu_{\rk}$ is real we have $(Kv|u)= \overline{(u| Kv)}$, hence using \eqref{e1.2z} twice we obtain:
\[
\bea
\int_{\Omega^+}(\bar{v}Ku+ \bar{Kv}u)\diff\mu_{\rk}&= \int_{\Omega^+}(\nabla_{a}^{(\rk)}\bar{v}\rk^{ab}\nabla_{b}^{(\rk)}u + \nabla_{b}^{(\ovk)}\bar{v}\ovk^{ab}\nabla_{a}^{(\ovk)}u + {( \lambda+\bar\lambda)} \bar{v}u) \diff\mu_{\rk}\\[2mm]
&\phantom{=\,} -\int_{\p\Omega}( \overline{n^{a}\nabla_{a}^{(\rk)}v}u+ \bar{v}n^{a}\nabla_{a}^{(\rk)}u )\diff\sigma_{\rh},
\eea
\]
which yields the desired result. \qed


If $F(X)\subset \cD'(X)$ is a space of distributions, we denote by $\bar{F}(\Omega^\pm)\subset \cD'(\Omega^\pm)$ the space of restriction{s} of elements of $F(X)$ to $\Omega^\pm$. This way, for instance, $\bar{\ccf}(\Omega^\pm)$ is the space of smooth functions on $\Omega^\pm$, smoothly extendible across $\p\Omega$, and with compact support in $X$. We will also frequently use the spaces $\bar{H^s_\cs}(\Omega^\pm)$ and $\bar{H^s_\loc}(\Omega^\pm)$ for $s\in\rr$, which are obtained by restricting elements of the compactly supported and local Sobolev spaces $H^s_\cs(X)$ and $H^s_\loc(X)$.

We set for $u\in \overline{C^{\infty}}(\Omega^{\pm})$:

\[
\gamma^{\pm}u\defeq\col{u\tra{\pOmega}}{n^{a}\nabla_{a}^{(\rk)}\!u\tra{\pOmega}},
\]
where the difference between the definition of $\gamma^+$ and $\gamma^-$ is that the trace is taken from respectively $\Omega^{+}$ and $\Omega^-$. We denote by $(\cdot| \cdot)_{\Omega^{\pm}}$ the scalar product induced from $(\cdot|\cdot)$ on $\Omega^\pm$.

\begin{proposition}
 \label{p1.1z}
 Let $u\in \overline{H^1_\cs}(\Omega^{\pm})$ and $v\in \overline{H^1_\loc}(\Omega^{\pm})$, or $u\in \overline{H^1_\loc}(\Omega^{\pm})$ and $v\in \overline{H^1_\cs}(\Omega^{\pm})$. Then:
 \[
\begin{array}{rl}
i)&(v| Ku)_{\Omega^{\pm}}- (K^{*}v| u)_{\Omega^{\pm}}= \pm(\gamma^{\pm}v| S \gamma^{\pm}u)_{\pOmega},\\[2mm]
ii)&(v| Ku)_{\Omega^{\pm}}+ (Kv|u)_{\Omega^{\pm}}= \eta^\pm(v,u)\mp (\gamma^{\pm}v| q \gamma^{\pm}u)_{\pOmega},
\end{array}
\]
where
\beq\label{eq:eta}
\eta^\pm(v,u)=2\int_{\Omega^{\pm}}(\nabla_{a}^{(\rk)}\bar{v}\Re \rk^{ab}\nabla^{(\rk)}_{b}u+(\Re\lambda) \bar{v}u) \diff\mu_{\rk},
\eeq
and setting $b= \Im n^{a}\nabla_{a}^{(\rk)}$,
\beq\label{eq:qS}
q= \mat{0}{\one}{\one}{0}, \ S= \mat{2\i b^{*}}{-\one}{\one}{0}, 
\eeq
where $b^{*}$ is the formal adjoint of $b$ in $L^{2}(\pOmega, d\sigma_{\rh})$.
\end{proposition}
\proof 
We obtain {\it ii)} for $\Omega^{+}$ by extending \eqref{e1.4z} by continuity,  the proof for $\Omega^{-}$ is similar.  To prove {\it i)} for $\Omega^{+}$ we rewrite \eqref{e1.1z} as:
\[
(v|Ku)_{\Omega^{+}}- (K^{*}v| u)_{\Omega^{+}}= (\bar{n}^a \nabla_{a}^{(\ovk)} v| u)_{\pOmega}- (v| n^a \nabla_{a}^{(\rk)} u)_{\pOmega}.
\]
Using the identity $\bar{n}^a \nabla_{a}^{(\ovk)}= n^a \nabla_{a}^{(\rk)} - 2 \i b$ concludes the proof {for $\Omega^+$. The $\Omega^-$ case is analogous.} \qed 

\subsection{Inverses of $K$}

In order to define what we will call \emph{Calder\'on operators}, we first need to have inverses of the operators $K$ and $K^*$ in an appropriate sense. We choose here to work  in a general abstract framework rather than focus on a case-by-case analysis.

We write $A\Subset B$ if $A$ is relatively compact in $B$.


\begin{hypothesis}\label{hyp:K} \textbf{\emph{(i)}} We assume that there exists  a continuous linear operator
\beq\label{eq:Kiso}
K^{-1} : H^{-1}_{\rm c}(X) \to H^{1}_{\loc}(X)
\eeq
such that $KK^{-1}=\one$ on $H^{-1}_{\rm c}(X)$ and $K^{-1} K =\one$ on $H^1_{\rm c}(X)$. \smallskip

\noindent\textbf{\emph{(ii)}} For all $v \in \overline{H^1_\cs}(\Omega^{\pm})$, $\eta^\pm(v,v)\geq 0$, where $\eta^\pm$ is defined in \eqref{eq:eta}.

\smallskip

\noindent\textbf{\emph{(iii)}} There exists a sequence $(\chi_i)$, $\chi_i\in\ccf(X;\rr)$, such that $\chi_i = \kappa \chi_i \kappa$ and:
\ben
\item[ \quad \quad a)] For any $Y\Subset X$, $\chi_i=1$ on $Y$ for $i$ sufficiently large.
\item[ \quad  \quad b)] For all $i$, {$\p_n \chi_i\traa\pOmega=0$.}
\item[ \quad  \quad  c)] For all $u,v\in \Ran K^{-1} \eqdef H^{1}_{\gl}(X)$, $(v|\chi_i [ K,\chi_i ]u) \to 0$ as $i\to\infty$. 
\een    
\end{hypothesis}


The subscript in $H^{1}_{\gl}(X)$ stands for `global', as the choice of inverse of $K$ (note that here we mean `inverse' in a somewhat weak way) in \eqref{eq:Kiso} is tied to the global behaviour. 
We remark that \textbf{{(iii)}} is trivially satisfied if $X$ is compact, as in that case one can  take $\chi_i=1$ for all $i$. 

\begin{example} The situation considered in \cite{analytic} is a special case of Hypothesis \ref{hyp:K} \textbf{{(i)}}, \noindent\textbf{{(ii)}} and \noindent\textbf{{(iii)}}. In \cite{analytic}, $\pOmega$ has one connected component (the Wick-rotated Cauchy surface), $X$ is embedded in a larger manifold, and the choice of inverse of $K$ corresponds to Dirichlet boundary conditions on the boundary of $X$.
\end{example}

\begin{example} An {intermediate} step in the construction of the Hartle-Hawking-Israel state in \cite{sanders,HHI,HHI2} consists in considering $K$ on a cylinder $X=\ss_\beta \times \Sigma$ for some smooth manifold $\Sigma$, where $\ss_\beta$ is the circle of length $\beta$. The choice $\Omega^+=\open{0,\beta/2}\times\Sigma$ and $\Omega^-=\open{-\beta/2,0}\times\Sigma$ gives two connected components of $\pOmega$ for each connected component of $\Sigma$. In the situation described in detail in \cite{HHI2}, $K$ defines a closable sesquilinear form on $\ccf(X)$, the closure of which is \emph{sectorial}. By the Lax-Milgram theorem, this defines an isomorphism from the form domain to its dual, which allows one to conclude that Hypothesis \ref{hyp:K} \textbf{{(i)}} is satisfied.
\end{example}

Thus, the scheme for constructing examples of \textbf{{(i)}} and \noindent\textbf{{(ii)}} is to study sectorial extensions of $K$ in the sense of sesquilinear forms, whereas  \textbf{{(iii)}} requires more specific knowledge of the domain. We expect that this analysis could be carried out basing on the recent works \cite{AGN,GN}. 

\begin{remark} In the case of complex metrics, the natural assumption for studying condition \textbf{{(i)}} through sectorial extensions of $K$ is the \emph{coercivity} of $\rk$, i.e., the existence of a continuous function  $C: X\to \open{0, +\infty}$ s.t.
 \beq\label{eq:coercivity}
|\Im (\bar{v}^{a}\rk_{ab}(x)v^{b})|\leq C(x) \Re (\bar{v}^{a}\rk_{ab}(x)v^{b}),  \ \forall x\in X, \ v\in \cc T_{x}X,
\eeq
see e.g.~\cite{HHI2} for the use of a \eqref{eq:coercivity} with $C$ uniform in $x$.
\end{remark} 

Let us now discuss several consequences of Hypothesis \ref{hyp:K}.  Since $K^*=\kappa \circ  K \circ  \kappa$ in the sense of formal adjoints, and the spaces $H^{{-1}}_{\rm c}(X)$, $H^{1}_{\loc}(X)$ are invariant under $\kappa$, from \eqref{eq:Kiso} we immediately conclude that $K^*$ has an inverse
\[
K^{*-1}\defeq \kappa\circ K^{-1} \circ  \kappa : H^{-1}_{\rm c}(X)\to H^{1}_{\loc}(X).
\]
By virtue of the next lemma, the definition of $K^{-1}$ and $K^{*-1}$ can  be extended to $H_\cs^s(X)$ for all $s<-1$, and $KK^{-1}=\one$ on $H_\cs^s(X)$ by continuity.

\begin{lemma}\label{l4.1}
 Let $\varphi_{1}, \varphi_{2}\in \coinf(X)$ with $\varphi_{1}= 1$ near $\supp \varphi_{2}$. Then there exists a classical, properly supported pseudo\-differential operator $Q$ {of order $-2$} such that the operator $\varphi_{1}K^{-1}\varphi_{2}- \varphi_{1}Q\varphi_{2}$ has smooth Schwartz kernel.
\end{lemma}

 Despite the setup  being more abstract here, the proof of \cite[Lem.~3.3]{analytic} generalizes in a straightforward way.

The two-sided trace operator is defined by
 \[
\gamma u= \col{u\traa{\p\Omega}}{\p_{n}u\traa{\p\Omega}}, \ \ u\in \cinf(X).
  \]
We denote by $\gamma^*$ its formal adjoint, defined using the two densities $d\mu_{\rk}$ and $d\sigma_{\rh}$. For $f\in C_{\rm c}^\infty(\p\Omega)^2$ we have $\gamma^* f \in H^{-2}_\cs(X)$, therefore $K^{-1} \gamma^* f$ is well defined at least as an element of $L^2_{\rm loc}(X)$. By repeating verbatim the proof of \cite[Lem.~A.1]{analytic} one gets the following more precise statement, where  we denote by $r^{\pm}: \cD'(X)\to \overline{\cD'}(\Omega^{\pm})$ the respective restriction operators.   

\begin{lemma}\label{lem:app1} The operator $r^{\pm}K^{-1} \gamma^*$ maps continuously $C_{\rm c}^\infty(\p\Omega)^2\to\overline{\cinf}(\Omega^{\pm})$, with range contained in $\overline{H^{1}_{\gl}}(\Omega^{\pm})$.
\end{lemma}

\subsection{Calder\'on operators and their properties} Recall that the operator $S$ was defined in \eqref{eq:qS}.

\begin{definition} We will call
 \[
C^{\pm}\defeq  \mp \gamma^{\pm}K^{-1} \gamma^{*}S 
\]
the \emph{\calde operators} associated with $K^{-1}$. 
\end{definition}

By Lemma \ref{lem:app1}, $C^\pm:\coinf(\pOmega)^{2}\to\cinf(\pOmega)^{2}$ continuously.

 If $X$ is compact then $C^\pm$ are the \emph{\calde projectors}. However, if $X$ is not compact then $C^\pm$ are not necessarily projections even if $\p\Omega$ is compact. This can arise for instance if $X=I\times\p\Omega$ with $I$ an \emph{open} interval. An example of such situation will be studied in Section \ref{sec:kms} in the context of \emph{KMS states}, cf.~\cite[12.9.3]{gerardbook} for explicit formulae in which the Calder\'on operators are components of a projection (in this case, this projection is itself a Calder\'on projector for an enlarged problem {where the boundary has an extra component}).  

\medskip

In what follows, by slight abuse of notation, if $\chi\in\ccf(X)$ and {$\p_n \chi\traa\pOmega=0$} then we simply write $\chi$ instead of $\chi\tra{\pOmega}$. 

\begin{lemma}\label{lem:long}  For any $\chi\in\ccf(X;\rr)$ such that {$\p_n \chi\traa\pOmega=0$}, and any $f,g\in \ccf(\pOmega)^2$, we have:
\beq\label{eq:lC1}
(\chi C^\pm f| q \chi C^\pm f )_{\pOmega} = \pm \eta^\pm(\chi u^\pm,\chi u^\pm) \mp 2 \Re (u^\pm | \chi [K,\chi]   u^\pm)_{\Omega^\pm},  
\eeq
\beq\label{eq:lC2}
(\chi C^- g | q \chi C^+ f )_{\pOmega} = ( w^{+}|\tilde\chi [K,\chi] u^{+})_{\Omega^{+}}- (\chi [K^{*},\tilde\chi] w^{+}| u^{+})_{\Omega^{+}},  
\eeq
where $u^{\pm}= \mp r^{\pm}K^{-1}\gamma^{*}Sf$, $w^{\pm}=\mp r^{\pm}\is K^{-1}\gamma^{*}Sg$ and $\tilde\chi=\is \chi \is$.  
\end{lemma}
\proof We apply Prop.~\ref{p1.1z} {\it ii)} to $u=v=\chi u^{\pm}$ as above. We obtain that:
\[
2\Re (\chi u^{\pm}| K \chi u^{\pm})_{\Omega^{\pm}}= \eta^\pm(\chi u^\pm,\chi u^\pm) \mp (\gamma^{\pm}\chi u^{\pm}| q \gamma^{\pm}\chi u^{\pm})_{\pOmega}.
\]
Since $K  u^{\pm}=0$ on $\Omega^\pm$ we can replace $K \chi u^{\pm}$ by $[K,\chi] u^{\pm}$. In view of $\gamma^{\pm}\chi u^{\pm}= \chi C^{\pm}f$ this implies \eqref{eq:lC1}.

Next, by Prop.~\ref{p1.1z} {\it i)} applied to $u=\chi u^+$ and $v=\tilde\chi w^+$ we obtain:
\beq\label{e1.10z}
(\tilde\chi w^{+}| K \chi u^{+})_{\Omega^{+}}- (K^{*}\tilde\chi w^{+}| \chi u^{+})_{\Omega^{+}}= (\gamma^{+}\tilde\chi w^{+}| S \gamma^{+}\chi u^{+})_{\pOmega}.
\eeq
We have $Ku^{+}= 0$ in $\Omega^{+}$ and by $K^*=\is K \is$  we have $K^{*}w^{+}= 0$ in $\Omega^{+}$.  Therefore we can rewrite \eqref{e1.10z} as
\beq\label{e1.10zz}
( w^{+}|\tilde\chi [K,\chi] u^{+})_{\Omega^{+}}- (\chi [K^{*},\tilde\chi] w^{+}| u^{+})_{\Omega^{+}}= (\gamma^{+}\tilde\chi w^{+}| S \gamma^{+}\chi u^{+})_{\pOmega}.
\eeq
 By 1) of \eqref{e1.8z} we have
\[
 (n^a \nabla_{a}^{(\rk)} \circ \is)\tra{\pOmega}= - (\bar{n}^a \nabla_{a}^{(\bar\rk)} )\tra{\pOmega}.
\]
Therefore,
\begin{equation}
\label{e1.9z}
\gamma^{+} u\circ \is= I \gamma^{-}u, \ \  I= \mat{\one}{0}{2\i b}{- \one},
\end{equation}
where we recall that $b= \Im n^a \nabla_{a}^{(\rk)} $. Thus,
 \[
(\gamma^{+}\tilde\chi w^{+}| S \gamma^{+}\chi u^{+})_{\pOmega}= -( I \chi C^{-}g| S \chi C^{+}f)_{\pOmega}= -(\chi C^{-}g| I^{*} S \chi C^{+}f)_{\pOmega}.
\]
Noting that 
\[
I^{*}S=\mat{\one}{-2 \i b^{*}}{0}{-\one} \mat{2\i b^{*}}{-\one}{\one}{0}=\mat{0}{-\one}{-\one}{0}= -q,
\] 
we obtain
\beq\label{eq:sd}
(\gamma^{+}\tilde\chi w^{+}| S \gamma^{+}\chi u^{+})_{\pOmega}=(\chi C^{-}g| q \chi C^{+}f)_{\pOmega}.
\eeq
Comparing \eqref{eq:sd}  with \eqref{e1.10zz} yields \eqref{eq:lC2}.
\qed

\begin{theoreme}\label{thm:calderon} Assume Hypothesis \ref{hyp:coercive} (ellipticity hypothesis), Hypothesis \ref{hyp:kappa} (existence of isometric involution $\kappa$), and Hypothesis \ref{hyp:K} \textbf{\emph{(i)}} (existence of $K^{-1}$) and  \textbf{\emph{(ii)}} (sufficient non-negativity of mass). Then the \calde operators $C^\pm$ are well-defined and satisfy
\beq\label{e1.5z}
C^{+}+ C^{-}= \one \mbox{ on } \ccf(\pOmega)^2.
\eeq
If moreover, the global condition Hypothesis \ref{hyp:K} \textbf{\emph{(iii)}} is satisfied, then
\beq\label{eq:positivity}
\pm (f|q C^\pm f)\geq 0,  \ \forall f \in \ccf(\pOmega)^2. 
\eeq
\end{theoreme}
\proof 
Let $f, g\in \coinf(\pOmega)^{2}$, $u^{\pm}= \mp r^{\pm}K^{-1}\gamma^{*}Sf$ and $v\in \coinf(X)$ such that $\gamma v= g$.
We have:
\[
\bea
(g| S(C^{+}+ C^{-})f)_{\pOmega}&= (\gamma^{+}r^{+}v| SC^{+}f)_{\pOmega}+ (\gamma^{-}r^{-}v| SC^{-}f)_{\pOmega}\\[2mm]
&=(r^{+}v| K u^{+})_{\Omega^{+}}- (r^{+}K^{*}v| u^{+})_{\Omega^{+}}
\\ &\phantom{=}\, - (r^{-}v| K u^{-})_{\Omega^{-}}+ (r^{-}K^{*}v| u^{-})_{\Omega^{-}}\\[2mm]
&=- (r^{+}K^{*}v| u^{+})_{\Omega^{+}}+ (r^{-}K^{*}v| u^{-})_{\Omega^{-}}= (K^{*}v| K^{-1}\gamma^{*}Sf)\\[2mm]
&=(v| \gamma^{*}Sf)= (\gamma v| Sf)_{\pOmega}= (g| Sf)_{\pOmega}.
\eea
\]
In line 1 we use that $\gamma v= \gamma^{\pm}r^\pm v$ since $v\in \coinf(X)$. In line 2 we use Prop.~\ref{p1.1z} {\it i)}. Then we use that $Ku^{\pm}= 0$ in $\Omega^{\pm}$ and the definition of $u^{\pm}$.
Since $S$ is invertible with $S^{-1}= \mat{0}{\one}{-\one}{2\i b^{*}}$ this implies \eqref{e1.5z}.

Next, let $\chi_n$ be as in Hypothesis \ref{hyp:K}. For $n$ large enough,
 \beq\label{eq:slkn}
 \bea
 (f| q C^{+}f)_{\pOmega}&= (\chi_{n}^{2}f|q C^{+}f)_{\pOmega}= (\chi_{n}f| q \chi_{n}C^{+}f)_{\pOmega}\\
 & = (\chi_{n}C^{+}f| q \chi_{n}C^{+}f)_{\pOmega}+ (\chi_{n}C^{-}f| q \chi_{n}C^{+}f)_{\pOmega}\\
 & =  \eta^+(\chi_n u^+,\chi_n u^+)  + c_n  
 \eea
 \eeq
 where by Lemma \ref{lem:long},
\[
\bea
c_n &=  - 2 \Re (u^+ | \chi_n [K,\chi_n]   u^+)_{\Omega^{+}}+ ( w^{+}|\chi_n [K,\chi_n] u^{+})_{\Omega^{+}}
\\ &\phantom{=} \, - (\chi_n [K^{*},\chi_n] w^{+}| u^{+})_{\Omega^{+}},  
\eea
\]
where $u^+$ and $w^+=- r^{+}\is K^{-1}\gamma^{*}S{f}$ belong to $\overline{H^{1}_{\gl}}(\Omega^{+})$ by Lemma \ref{lem:app1}. By Hypothesis \ref{hyp:K} \textbf{{(iii)}}, $c_n\to 0$ as $n\to+\infty$. By Hypothesis \ref{hyp:K} \textbf{{(ii)}}, the $\eta^+$ term in \eqref{eq:slkn} is positive, hence  $(f| q C^{+}f)_{\pOmega}\geq c_n$. Thus, letting $n\to +\infty$ we obtain the `$+$' version of \eqref{eq:positivity}. The `$-$' variant is proved analogously.   \qed

We end this section with a simple auxiliary lemma, thanks to which the positivity statement \eqref{eq:positivity} can be interpreted as `reflection positivity' of $K^{-1}$ at $\p\Omega$. Note that positivity of $q C^+$ is equivalent to the positivity of $q^* (q C^+) q=C^+ q$. 

\begin{lemma}\label{lem:S0} We have $C^+ q = \gamma^+ K^{-1} \is \gamma^*$.
\end{lemma}
\proof Recall that the operator $I$ was defined in \eqref{e1.9z}. We have
\beq\label{eq:tmss}
S^{-1}I^* =\begin{pmatrix} 0 & \one \\ -\one & 2\i b^* \end{pmatrix} \begin{pmatrix}\one & - 2\i b^* \\ 0 & -\one \end{pmatrix} = \begin{pmatrix} 0 & -\one \\ -\one & 0 \end{pmatrix} = -q.
\eeq
Using the identity $\is \gamma^* = \gamma^* I^*$ and   \eqref{eq:tmss} we obtain
\beq\label{eq:slkm}
\bea
\gamma^+ K^{-1} \is \gamma^* & = \gamma^+ K^{-1}  \gamma^* I^* = \gamma^+ K^{-1}  \gamma^* S S^{-1} I^{*} \\
 & = -\gamma^+ K^{-1} \gamma^* S q =  C^+ q
\eea
\eeq
as claimed. \qed

\section{Two-point functions from Calder\'on operators}\label{sec:wick}\init

\subsection{Geometric setup}\label{ss:geometric} 

 We now consider the following setup, in which $\pOmega$ has only one connected component:

\begin{enumerate}[label={\arabic*)}]
\item Let $\Sigma$ be a connected smooth manifold. \smallskip
\item Let $\rh : \cc \to \cc {\rm T}^{0}_{2}(\Sigma)$, $N,\mu: \cc \to \cc {\rm T}^{0}_{0}(\Sigma)$, and $w:\cc\to\cc {\rm T}^{1}_{0}(\Sigma)$ be complex tensor valued functions such that
\[
\overline{\rh(z)}=\rh(\bar{z}), \ \ \overline{N(z)}=N(\bar{z}), \ \ \overline{\mu(z)}=\mu(\bar{z}), \ \ \overline{w^{i}(z)}=w^i(\bar{z}), \ \ z\in \cc.
\]
\item Writing $z=t+\i s$, we assume that
\[
\bea
\rg&\defeq - N^{2}(t) dt^{2}+ \rh_{jk}(t)(dy^{j}+ w^{j}(t)dt)(dy^{k}+w^{k}(t)dt),\\
\rk&\defeq N^{2}(\i s)ds^{2}+ \rh_{jk}(\i s)(dy^{j}+ \i w^{j}(\i s)d s)(dy^{k}+ \i w^{k}(\i s)d s),
\eea
\]
are complex metrics on $\rr\times\Sigma$, {$\rg$ is Lorentzian}, and $\rk$ satisfies Hypothesis \ref{hyp:coercive}. We assume that $\mu(t,y)$ and $\lambda(s,y)\defeq\mu(\i s,y)$ are smooth. \smallskip
\item Let $\kappa:\rr\times\Sigma\to \rr\times\Sigma$ be the diffeomorphism
\[
\kappa(s,y)\defeq (-s,y). 
\]
\item Let $\Me$ be a $\kappa$-invariant neighborhood of $\{0\}\times\Sigma$. We set:
 \[
\Omega^{\pm}\defeq X\cap\{ \pm s>0\}, \ \ \p\Omega=\{s=0\}.
\]
\end{enumerate}\smallskip

\noindent By 2), the metric $\rg$ introduced in $3)$ is real. The metric $\rk$ is the Wick-rotated counterpart of $\rg$, and does not need to be real. 

\medskip

The definitions 1)--5) match the setup of Section \ref{sec:cald}, with Hypothesis \ref{hyp:kappa} satisfied. Recall that we have defined there
\[
 K=  - |\rk|^{-\12}\p_{a}\rk^{ab}|\rk|^{\12}\p_{b}+\lambda.
\]
If in addition, we have an inverse of $K$ in the sense of Hypothesis \ref{hyp:K}, then all the results of Section \ref{sec:cald} apply and we have a pair of \calde operators $C^\pm$.  

 Our main object of interest is the Klein-Gordon operator $P$ for the Lorentzian metric $\rg$, namely:

\begin{enumerate}
\item[6)] Let $(M,\rg)$ be the globally hyperbolic spacetime defined as the domain of dependence of $\{0\}\times\Sigma$. \smallskip
\item[7)] On $M$ we consider the Klein-Gordon operator 
\[
P= -|\rg|^{-\12}\p_{a}|\rg|^{\12}\rg^{ab}\p_{b}+\mu.
\]
\end{enumerate}

\begin{remark} We remark that Chru\'sciel and Delay have derived a method to obtain non-singular real-analytic solutions of Einstein equations of the form 3), though they applied it to the negative cosmological constant case \cite{CD}. Here we limit ourselves to globally hyperbolic regions due to the use of analytic propagation of singularities theorems in the proof of Prop.~\ref{reform}.  
\end{remark}

\subsection{Two-point functions}\label{ss:tpf} Let $n\in TM$ be the outer unit normal vector field to $\{0\}\times\Sigma$. For $f\in \cD'(\Sigma)^{2}$ we denote by $Uf\in \cD'(M)$ the unique solution of $Pu=0$ with Cauchy data at $t=0$ given by
\[
\begin{pmatrix} u\traa{\Sigma} \\  \i^{-1}\p_n u\traa{\Sigma} \end{pmatrix}=f.
\]
Let $G$ be the difference of the retarded and advanced propagators for $P$. It is well known that
\beq\label{eq:Uooo}
 U q  U^*= \i G,
\eeq
where $U^*$ is the formal adjoint of $U$ w.r.t.~the canonical $L^2(M,\rg)$ scalar product.

\medskip

As a direct corollary of Theorem \ref{thm:calderon} and \eqref{eq:Uooo}, in the setup of definitions 1)--7) we have:

\begin{proposition}\label{prop2pt} Let $C^\pm$ be the \calde operators associated with an inverse of $K$ satisfying Hypothesis \ref{hyp:K}.  Set
\beq
\Lambda^\pm\defeq {\pm} U C^{\pm}q  U^* : \ccf(M)\to\cf(M).
\eeq
Then $\Lambda^\pm$ is a pair of two-point functions, i.e.
\beq\label{eq:podo}
\Lambda^\pm\geq 0, \  P\Lambda^\pm=\Lambda^\pm P=0, \mbox{ and } \Lambda^+-\Lambda^-=\i G \mbox { on } \ccf(M).
\eeq
\end{proposition}

We refer the reader to e.g.~\cite{analytic} for an explanation of the importance of the conditions \eqref{eq:podo} in Quantum Field Theory.

\section{Wick rotation in the analytic case}\label{sec:anal}

\subsection{Geometric setup} From now on we consider the same setup as in Subsection \ref{ss:geometric}, but supposing in addition that $\Sigma$ is a \emph{real analytic} manifold and $\rg$ and $\mu$ are \emph{analytic}. 

\subsection{Boundary values of holomorphic functions} Let us introduce the conventions and basic notions used throughout this section.

\begin{notations}
\item 
If $\Gamma, \Gamma'$ are two cones of vertex $0$ in $\rr^{n}$ we write $\Gamma'\Subset \Gamma$ if $(\Gamma'\cap\mathbb{S}^{n-1})\Subset (\Gamma\cap \mathbb{S}^{n-1})$.
\item Let $\Gamma$ be an arbitrary convex, open and proper cone of vertex $0$ in $\rr^{n}$, or in short, an \emph{open cone}. Its \emph{polar} is then defined by
\[
\Gamma^{\circ}\defeq \{\xi\in \rr^{n}: \ \xi\cdot y\geq 0, \ \ \forall y\in \Gamma\}.
\]
\item Let $\Omega\subset \rr^{n}$ be open. Then a {\em tuboid of profile} $\Omega+ \i \Gamma$ is a complex domain $D\subset \Omega+ \i \Gamma$ such that for all $x\in \Omega$ and  any  subcone $\Gamma'\Subset \Gamma$ there exists  a neighborhood $\Omega'$ of $x$ in $\Omega$ and a constant $r>0$ such that 
\beq\label{tototo}
\Omega'+ \i \{y\in \Gamma': \ 0<|y|\leq r\}\subset D.
\eeq
\item We write  $F\in \mo_{\rm temp}(\Omega+ \i \Gamma 0)$ if $F$ is a holomorphic function on some tuboid $D$ of profile $\Omega+ \i \Gamma$, and $F$ is {\em temperate}, i.e.~for any $K\Subset \Omega$, any  subcone $\Gamma'\Subset \Gamma$ and any $r>0$ such that $K+\i\{y\in  \Gamma': \ 0<|y|\leq r\}\subset D$, there exists $C, r'>0$ and $N\in \nn$ such that
\[
|F(x+ \i y)|\leq C |y|^{-N}, \ \ x\in K, \  y\in \Gamma', \ 0<|y|\leq r'.
\]
\end{notations}

\medskip

\noindent The importance of temperate holomorphic functions is that their boundary values are distributions. This is stated precisely in the next theorem (see e.g.~\cite[Thm.~3.6]{Kom} for the proof).

\begin{theoreme}\label{thm:equi} Let $F$ be a holomorphic function on a tuboid $D$ of profile $\Omega+\i \Gamma$. Then the following are equivalent:
\begin{enumerate}[label={\alph*)}]
\item $F\in \mo_{\rm temp}(\Omega+ \i \Gamma 0)$;
\item For all $x\in \Omega$ and some closed cone $\Gamma'\Subset\Gamma$, there exists a neighbourhood $\Omega'$ of $x$ and $r>0$ such that \eqref{tototo} holds true and the set $\{ F(x+\i y)  : \ y\in\Gamma', \ |y|< r\}$ is bounded in $\cD'(\Omega')$;
\item For all $x\in \Omega$ and all closed cones $\Gamma'\Subset\Gamma$, there exists a neighbourhood $\Omega'$ of $x$ and $r>0$ such that \eqref{tototo} holds true and the set $\{ F(x+\i y)  : \ y\in\Gamma', \ |y|< r\}$ is bounded in $\cD'(\Omega')$;
\item The limit 
\[
F(x+\i\Gamma 0)\defeq \lim_{\Gamma'\ni y\to 0}F(x+ \i y)
\]
exists in $\cD'(\Omega)$ for any $\Gamma'\Subset \Gamma$ (in the precise sense of \cite[Def.~3.4]{Kom}).
\end{enumerate}
\end{theoreme}

\noindent We will need a generalization to distribution-valued holomorphic functions. Let $Z$ be a smooth manifold, and let us denote by $\langle \cdot, \cdot \rangle$ the duality bracket between $\cD'(Z)$ and $\ccf(Z)$.

\begin{notations}
\item We write $F\in\mo_{\rm temp}(\Omega + \i \Gamma 0; \cD'(Z))$ if $F$ is a $\cD'(Z)$-valued holomorphic function on some tuboid $D$ of profile $\Omega+ \i \Gamma$, and for each  $K\Subset \Omega$  there exist $r>0, N\in \nn$ such that  for each  bounded set $\cB\subset \ccf(Z)$, there exists $C_{\cB}>0$ such that
 \[
 \sup_{\varphi\in \cB}|\langle F(x+\i y, \cdot), \varphi(\cdot)\rangle|\leq C_{\cB}|y|^{-N}, \ \ x\in K, \ 0<|y|\leq r. 
 \]
\item By convention, $\Omega - \i \Gamma 0= \Omega + \i (-\Gamma) 0$.
\item If $I\subset \rr$ is an open interval and $\Gamma=\open{0,+\infty}$ then we simply write $I\pm \i 0$ in the place of $I\pm \i \Gamma$.
\end{notations}

\medskip

\noindent By Theorem \ref{thm:equi}, if $F\in\mo_{\rm temp}(\Omega \pm \i \Gamma 0; \cD'(Z))$ then
\beq\label{defigammao}
\langle F(x\pm\i\Gamma 0,\cdot), \varphi(\cdot) \rangle \defeq \lim_{\Gamma'\ni y\to 0} \langle  F(x\pm \i y,\cdot),\varphi(\cdot) \rangle
\eeq
defines uniquely a distribution in $\cD'(\Omega\times Z)$.

\subsection{Analytic Hadamard condition}

In \cite[Prop.~5.2]{analytic} it was shown (in a less general context) that solutions of $Pu=0$ with Cauchy data in the range of $C^+$ are boundary values of holomorphic functions from $\{ s> 0\}$. The crucial observation for us is that if we view this as an analytic continuation result for solutions of $Kv=w$ with $w$ supported on $\p\Omega=\Sigma$, then the proof extends to much more general situations. The trick is not to try to study solutions of $Kv=w$ directly on $X$, but instead restrict $v$ to $\Omega^+$ and then extend it by zero to $X$.

If $I\subset \rr$, we set $I^\pm = I\cap\{\pm s >0\}$. For $Z$ a smooth manifold, we will write $F\in\mo_{\rm temp}(I^+\times \i I^+;\cD'(Z))$ if $F\in\mo_{\rm temp}(\rr+\i 0;\cD'(Z))$ and $F$ is holomorphic in $I^+\times \i I^+$ (rather than merely on a subdomain).

\begin{proposition}\label{reform} Suppose that $\Sigma$ is a real analytic manifold and $\rg$ and $\mu$ are analytic. Let $v\in\overline{\cf}(\Omega^+;\cD'(Z))$ for $Z$ a smooth manifold and suppose that
\[
\left( (K\otimes \one_Z) v\right)(s,y,z)=0 \  \mbox{ in a neighborhood of } \{0\}\times\Sigma\times Z \mbox{ in } \Omega^+\times Z. 
\]
Let $u=(U \gamma^+ \otimes \one_Z) v \in \cD'(M\times Z)$. Then for any $y^0\in\Sigma$ there exists a neighborhood $Y\subset \Sigma$, an interval $I=\open{-\delta,\delta}$, and a distribution-valued function $F\in\mo_{\rm temp}(I^+\times \i I^+;\cD'(Y\times Z))$ such that:
\beq
\bea
v(s,y,z)&=F(\i s,y,z), \ \ s\in I^{+},\\
u(t,y,z)&=F(t+\i 0,y,z), \ \ t\in I.
\eea
\eeq  
\end{proposition}
\proof Let $v_{\rm ext}=(e^+\otimes\one_Z) v$, where $e^+:\overline{\cf}(\Omega^+)\to\cD'(X)$ is the distributional extension by zero. Then inside of a neighborhood of $\{0\}\times\Sigma\times Z$,  $(K\otimes \one_Z)v_{\rm ext}$ is supported in $\{s=0\}$. We can now repeat the steps in the proof of Prop.~5.2 in \cite{analytic} to get the desired result. To explain this very briefly, one writes $v_{\rm ext}(s)$ as the difference $v^{\rm r}_{\rm ext}(s)-v^{\rm l}_{\rm ext}(s)$ of two boundary values $s{\mp} \i 0$. Wick-rotating $v^{\rm r}_{\rm ext}$ and $v^{\rm l}_{\rm ext}$ yields two distributions $u^{\rm r}(t)$ and $u^{\rm l}(t)$ defined on $\{\pm t>0\}$, which solve $P u^{\rm r/\rm l}(t) = w(t+\i 0)$ for a suitable temperate $w$ (in general different from the one considered in \cite{analytic}) obtained by Wick-rotating $K v^{\rm r/\rm l}_{\rm ext}$. One can extend $u^{\rm r/\rm l}$ to all $t\in\rr$ using the hyperbolicity of $P$, and then one represents the resulting distribution as a boundary value of some temperate $\tilde{v}^{\rm r/\rm l}_{\rm ext}$, which is possible by analytic wave front set arguments. The difference $F=\tilde{v}^{\rm r}_{\rm ext}-\tilde{v}^{\rm l}_{\rm ext}$ satisfies the stated properties. \qed

In the same way as in Prop.~5.3 in \cite{analytic} we can also show:

\begin{proposition}\label{cor:had} Let $\Lambda^\pm$ be as in Prop.~\ref{prop2pt}. Suppose that $\Sigma$ is a real analytic manifold and $\rg$ and $\mu$ are analytic.  Then the pair $\Lambda^\pm$ satisfies the \emph{analytic Hadamard condition} in $M$, i.e.
\beq\label{eq:hada}
\WFA(\Lambda^\pm)\subset \cN^{\pm}\times \cN^{\mp},
\eeq
where, denoting by  $V_{x\pm}$ the future/past (solid) lightcones in $T_x M$ and by $\zero$ the zero section of $T^*M$,
\[
\cN^\pm = \{(x,\xi)\in T^*M\setminus\zero : \ \xi\cdot \rg^{-1}(x)\xi =0 \mbox{ and } \xi\cdot v  > 0 \ \forall v\in V_{x\pm} \mbox{ s.t. } v\neq 0\}.
\]
\end{proposition}

The analytic Hadamard condition was introduced by Strohmaier, Verch and Wollenberg \cite{SVW}, see \cite{analytic} for the equivalence between the original formulation and the one above, inspired by \cite{SV}. In \eqref{eq:hada}, $\WFA(\Lambda^\pm)$ stands for the \emph{analytic \wavefront set} of the Schwartz kernel $\Lambda^\pm(t_1,y_1,t_2,y_2)\in\cD'(M^2)$ of $\Lambda^\pm$. Instead of recalling the precise definition (see e.g.~\cite{H3,Kom}), we prefer to indicate the consequences of \eqref{eq:hada} which will be directly relevant for our analysis. Focusing on $\Lambda^+$, we will actually only need the following apparently weaker inclusion:
\beq
\WFA(\Lambda^+)\subset V^{\circ}, \ \ V^{\circ}=\left\{ \tau_1\geq c |k_1|, \ \tau_2\leq-c|k_2|, \ c>0 \right\},
\eeq
where we wrote $(\tau_i,k_i)$ for the dual variables of $(t_i,y_i)$. Above, $V^{\circ}$ is the polar of $V= \{ s_1> c{^{-1}} |y_1|, \ s_2<-c{^{-1}}|y_2|\}$. This implies (see e.g.~\cite[Thm.~3.9]{Kom}) that there exists an interval $I$ and $F\in \mo_{\rm temp}((I\times \Sigma)^2+ \i V 0)$ such that 
\beq\label{eq:blabl}
\Lambda^+(t_1,y_1, t_2,y_2)= F((t_1,y_1,t_2,y_2)+ \i V0).
\eeq 
This statement will be the starting point for us.

In the next lemma we give a variant of \eqref{eq:blabl} in terms of holomorphic functions with values in Schwartz kernels on $\Sigma$. Let 
\[
\Gamma= \{s_1>0, \ s_2<0\} \subset \rr^2.
\]
Note that $\Gamma^{\circ}=\Gamma^{\rm cl}$. 

\begin{lemma}\label{lem:lambdaF} As a consequence of \eqref{eq:blabl}, there exists $F\in \mo_{\rm temp}(\rr^{2}+ \i\Gamma 0; \cD'(\Sigma^2))$ such that
\begin{equation}
\label{eq:lambdaF}
\Lambda^+(t_{1}, y_{1}, t_{2}, y_{2})= F((t_{1}, t_{2})+ \i\Gamma 0, y_{1}, y_{2}).
\end{equation}
\end{lemma}
\proof Recall that \eqref{eq:blabl} means that we have $F\in \mo_{\rm temp}((I\times \Sigma)^2+ \i V 0)$ such that
\[
\Lambda^+(t_{1}, y_{1}, t_{2}, y_{2})= \lim_{(\xi_1, \xi_2)\to 0}F\left((t_{1}, y_{1})+ \i \xi_1, (t_{2}, y_{2})+ \i \xi_2\right) \hbox{ in }\cD'((I\times \Sigma)^{2}),
\]
where the limit $(\xi_1, \xi_2)\to 0$ is taken from any closed cone $W\Subset V$. We stress that the notion of closed and compact refers here to the intersection of the various sets with the unit sphere. In particular we can take
\[
W= \{s_1>0, s_2<0, \ y_1= y_2= 0\}.
\]
Accordingly, we can view $F(z_1, z_2, y_1, y_2)$ as a $\cD'(\Sigma^2)$-valued function, holomorphic in $(z_1, z_2)\in I^{2}+ \i \Gamma 0$.  By Theorem \ref{thm:equi}, $F(t_1+ \i s_1, t_2+ \i s_2, y_1, y_2)$ is bounded in $\cD'((I\times \Sigma)^{2})$ when $(s_1, s_2)\in \Gamma$.  This means that if  $\cB$  is a bounded set in $\ccf(\Sigma^{2})$  then 
\[
F_{\varphi}(t_1+ \i s_1, t_2+ \i s_2)\defeq \langle F(t_1+ \i s_1, t_2+ \i s_2,\cdot, \cdot), \varphi\rangle
\]
is bounded in $\cD'(I^{2})$ for $(s_1,s_2)\in \Gamma$ and $\varphi\in \cB$.  Equivalently, by Theorem \ref{thm:equi} there exists  $N\in \nn$ and $C>0$ such that
\[
|F_{\varphi}(t_1+ \i s_1, t_2+ \i s_2)|\leq C |s_1|^{-N}|s_2|^{-N}, \ \varphi\in \cB,  \ (s_1, s_2)\in \Gamma.
\]
Furthermore, for $\psi\in \ccf(I^{2})$ we have
 \[
\langle F_{\varphi}((t_1,t_2)+ \i \Gamma 0), \psi\rangle= \langle F((x_1, x_2)+ \i V0), \psi\otimes \varphi\rangle,
\]
hence:
\[
\Lambda^+(t_1,  y_1, t_2,y_2)= F((t_1, t_2)+ \i \Gamma 0, y_1, y_2),
\]
and $F(z_1, z_2, y_1, y_2)\in \mo_{\rm temp}(I^{2}+ \i \Gamma 0; \cD'(\Sigma^{2}))$. \qed 

\subsection{Wick rotation in two time variables}

In what follows we will show that the Schwartz kernels of $K^{-1}$ and $\Lambda^+$ are related by Wick rotation in two time variables.

\begin{lemma}\label{lem:another} For {any} $y_1^0\in \Sigma$ there exists a neighborhood $Y_1$ of $y_1^0$, a neighborhood $U$ of $\{0\}\times\Sigma$, {an interval $I_1=]-\delta,\delta[$} and $F^+\in\mo_{\rm temp}(I^+_1\times \i I^+_1;\cD'(Y_1\times U))$ such that
\beq\label{eq:kk1}
\begin{array}{l}
\Lambda^+(t_1,y_1,t_2,y_2)=F^+(t_1+\i 0,y_1,t_2,y_2),\\[2mm]
  \hbox{ for } t_1\in I_1, \ y_{1}\in Y_{1}, \ (t_{2}, y_{2})\in U,
\end{array}
\eeq
\beq\label{eq:kk2}
\begin{array}{l}
\big(K^{-1}\is\gamma^*U^*\big)(s_1,y_1,t_2,y_2)= F^+(\i s_1,y_1,t_2,y_2), \\[2mm]
  \hbox{ for } s_1\in I_1^+, \ y_{1}\in Y_{1}, \ (t_{2}, y_{2})\in U.
\end{array}
\eeq
\end{lemma}
\proof By Lemma \ref{lem:app1} we can apply Prop.~\ref{reform} to the Schwartz kernel of $K^{-1} \is \gamma^* U^*$. For any $y_1^0\in \Sigma$, this gives the existence of an interval $I_1$,  a neighborhood $Y_1$ of $y_{1}^{0}$ in $\Sigma$, a neighborhood $U$ of $\{0\}\times \Sigma$ in $\rr\times \Sigma$ and $F^+\in\mo_{\rm temp}(I^+_1\times \i I^+_1;\cD'(Y_1\times U))$ such that
\begin{equation}
\label{e0.11}
\begin{array}{l}
\big(K^{-1} \is \gamma^* U^*\big)(s_1,y_1,t_2,y_2)= F^+(\i s_1,y_1,t_2,y_2), \\[2mm]
 \hbox{ for } s_1\in I^+_1, \ y_{1}\in Y_{1}, \ (t_{2}, y_{2})\in U,
\end{array}
\end{equation}
\begin{equation}
\label{e0.12}
\begin{array}{l}
\big(U \gamma^+ K^{-1} \is \gamma^* U^*\big)(t_1,y_1,t_2,y_2)=F^+(t_1+\i 0, y_1,t_2,y_2), \\[2mm]
  \hbox{ for } t_1\in I_1, \ y_{1}\in Y_{1}, \ (t_{2}, y_{2})\in U.
\end{array}
\end{equation} 
Using Lemma \ref{lem:S0} and the definition of $\Lambda^+=U C^+ q U^*$, we can rewrite \eqref{e0.12} as
\[
\begin{array}{l}
\Lambda^+(t_1,y_1,t_2,y_2)=F^+(t_1+\i 0, y_1,t_2,y_2), \\[2mm]
  \hbox{ for } t_1\in I_1, \ y_{1}\in Y_{1}, \ (t_{2}, y_{2})\in U
\end{array}
\]
as claimed. \qed

Our main result is the following theorem. In \eqref{eq:main} below, the limit from $\Gamma$ is meant as in \eqref{defigammao}, and the spatial variables $y_1,y_2$ are omitted in the notation. We remark that the theorem is still true if analyticity is assumed only for $t+\i s$ in a suitable neighborhood of $0$.

\begin{theoreme}\label{thm:main} Consider the setup from 1)--7)  in Subsect.~\ref{ss:geometric}. Assume that $\Sigma$ is a real analytic manifold, and that $\rg$ and $\mu$ are analytic. Suppose that $K^{-1}$ is an inverse of $K$ satisfying Hypothesis \ref{hyp:K}, and let $\Lambda^\pm$ be the associated pair of two-point functions defined in Prop.~\ref{prop2pt}. Let $\Gamma= \{s_1>0, \ s_2<0\}$. Then there exists an interval $I=\open{-\delta,\delta}$ and $F^\pm\in\mo_{\rm temp}(\rr^2\pm\i\Gamma 0;\cD'(\Sigma^2))$ such that:
\beq\label{eq:main}
\bea
K^{-1}(s_1,s_2)&= F^\pm(\i s_1, \i s_2), \ \ \pm s_1>0, \ {\pm} s_2<0,\\
\Lambda^\pm (t_1,t_2)&= F^\pm\big((t_1,t_2)\pm\i \Gamma 0\big), \ \ t_1,t_2\in I.
\eea
\eeq
\end{theoreme}
\proof We focus on the $+$ case, the $-$ case being analogous.

By ellipticity of $K$, the Schwartz kernel $K^{-1}(s_2,y_2,s_1,y_1)$ is in $\overline{\cf}(\Omega^+;\cD'(\Omega^-))$. For  $y_2^0\in\Sigma$, Prop.~\ref{reform} applied to $v=K^{-1} (s_2,y_2,s_1,y_1)$ and $Z=\Omega^-$ gives the existence of an interval $I_2$, a neighborhood $Y_2$ of $y_2^0$ and  $F_2\in\mo_{\rm temp}(I^+_2\times\i I^+_2;\cD'(Y_2\times \Omega^-))$ such that
\beq\label{eq:omg1} 
\begin{array}{l}
K^{-1} (s_2,y_2,s_1,y_1)= F_2(\i s_2,y_2,s_1,y_1), \\[2mm]
  \hbox{ for } s_2\in I^+_2, \ y_{2}\in Y_{2}, \ (s_{1}, y_{1})\in \Omega^{-},
\end{array} 
\eeq
\beq
\label{eq:omg2} 
\begin{array}{l}
 \left(U \gamma^+ K^{-1}\right) (t_2,y_2,s_1,y_1)=F_2(t_2+\i 0, y_2,s_1,y_1), \\[2mm]
  \hbox{ for } t_2\in I_2, \ y_{2}\in Y_{2}, \ (s_{1}, y_{1})\in\Omega^{-}.
\end{array}
\eeq
Taking formal adjoints of \eqref{eq:omg1}, we get
\beq\label{eq:omg3a}
(K^*)^{-1}(s_{1}, y_{1}, s_{2}, y_{2})= \overline{F_{2}(\i s_{2}, y_{2}, s_{1}, y_{1})}.
\eeq
Since $(K^*)^{-1}=\is  K^{-1}\is $, by composing both sides of \eqref{eq:omg3a} to the left with $\is$ we get:
\begin{equation}
\label{eq:omg3b}
\begin{array}{l}
\big( K^{-1} \is)(s_{1}, y_{1}, s_{2}, y_{2})=  \overline{F_{2}(\i s_{2}, y_{2}, \is(s_{1}, y_{1}))}, \\[2mm]
 \hbox{ for } (s_{1}, y_{1})\in\Omega^{+}, \ s_{2}\in I_{2}^{+},\ y_{2}\in Y_{2}.
\end{array}
\end{equation}
Taking formal adjoints of \eqref{eq:omg2} gives:
\begin{equation}
\label{eq: omg2bb}
\begin{array}{l}
\big((K^*)^{-1}\gamma^* U^*\big)(s_{1}, y_{1}, t_{2}, y_{2})= \overline{F_{2}(t_{2}+ \i 0, y_{2}, s_{1}, y_{1})},\\[2mm]
 \hbox{ for }  (s_{1}, y_{1})\in\Omega^{-}, \ t_{2}\in I_{2}, \ y_{2}\in Y_{2}.
\end{array}
\end{equation}
Setting $\overline{F}_{2}(z)\defeq \overline{F_{2}(\bar{z})}$ and using $(K^*)^{-1}=\is  K^{-1}\is $, we rewrite \eqref{eq:omg3b} and \eqref{eq: omg2bb} as:
\begin{equation}
\label{e0.7d}
\begin{array}{l}
\big(K^{-1}\is\big)(s_{1}, y_{1}, s_{2}, y_{2})= \bar{F}_{2}(-\i s_{2}, y_{2}, \is(s_{1}, y_{1})),\\[2mm]
 \hbox{ for } (s_{1}, y_{1})\in\Omega^{+}, \ s_{2}\in I_{2}^{+},\ y_{2}\in Y_{2}, 
\end{array}
\end{equation}
\begin{equation}
\label{e0.8c}
\begin{array}{l}
\big(K^{-1} \is \gamma^* U^*\big)(s_{1}, y_{1}, t_{2}, y_{2})= \bar{F}_{2}(t_{2}- \i 0, y_{2}, \is(s_{1}, y_{1})), \\[2mm]
 \hbox{ for }  (s_{1}, y_{1})\in\Omega^{+}, \ t_{2}\in I_{2}, \ y_{2}\in Y_{2}.
\end{array}
\end{equation}
Lemma \ref{lem:another} gives the existence of an interval $I_1$,  a neighborhood $Y_1$ of $y_{1}^{0}$ in $\Sigma$, a neighborhood $Z_2$ of $\{0\}\times \Sigma$ in $\rr\times \Sigma$ and $F_1\in\mo_{\rm temp}(I^+_1\times \i I^+_1;\cD'(Y_1\times Z_2))$ such that
\begin{equation}
\label{e0.11b}
\begin{array}{l}
\big(K^{-1} \is \gamma^* U^*\big)(s_1,y_1,t_2,y_2)= F_1(\i s_1,y_1,t_2,y_2), \\[2mm]
 \hbox{ for } s_1\in I^+_1, \ y_{1}\in Y_{1}, \ (t_{2}, y_{2})\in Z_2,
\end{array}
\end{equation}
\begin{equation}
\label{e0.12b}
\begin{array}{l}
\Lambda^+(t_1,y_1,t_2,y_2)=F_1(t_1+\i 0, y_1,t_2,y_2), \\[2mm]
  \hbox{ for } t_1\in I_1, \ y_{1}\in Y_{1}, \ (t_{2}, y_{2})\in Z_2.
\end{array}
\end{equation}
On the other hand, by Lemma \ref{lem:lambdaF} there exists $F\in \mo_{\rm temp}(\rr^{2}+ \i\Gamma 0; \cD'(\Sigma\times \Sigma))$ such that
\begin{equation}\label{eq:lambdaF2}
\Lambda^+(t_{1}, y_{1}, t_{2}, y_{2})= F((t_{1}, t_{2})+ \i\Gamma 0, y_{1}, y_{2}).
\end{equation}
By comparing \eqref{e0.12b} and \eqref{eq:lambdaF2}, after possibly shrinking $I_1$ and $I_2$,  we obtain that:
\begin{equation}
\label{e0.14}
\begin{array}{l}
F_{1}(z_{1}, y_{1}, t_{2}, y_{2})= F(z_{1}, t_{2}- \i 0, y_{1}, y_{2})\\[2mm]
 \hbox{ for } y_{i}\in Y_{i}, \ t_{2}\in I_{2}, \ z_{1}\in I_{1}+ \i I_{1}^{+}.
\end{array}
\end{equation}
By comparing \eqref{e0.8c} with \eqref{e0.11b} we obtain that:
\begin{equation}
\label{e0.15}
\begin{array}{l}
F_{1}(\i s_{1}, y_{1}, t_{2}, y_{2})= \bar{F}_{2}(t_{2}- \i 0, y_{2}, \is(s_{1}, y_{1}) )\\[2mm]
 \hbox{ for } y_{i}\in Y_{i}, \ s_{1}\in I_{1}^{+}, \ t_{2}\in I_{2},
 \end{array}
\end{equation}
hence by \eqref{e0.14}:
\[
\begin{array}{l}
\bar{F}_{2}(t_{2}- \i 0, y_{2}, \is( s_{1}, y_{1}))= F(\i s_{1}, t_{2}- \i 0, y_{1}, y_{2})\\[2mm]
\hbox{ for }y_{i}\in Y_{i}, \ s_{1}\in I_{1}^{+}, \ t_{2}\in I_{2},\\[2mm]
\bar{F}_{2}(z_{2}, y_{2}, \is(s_{1}, y_{1}))= F(\i s_{1}, z_{2}, y_{1}, y_{2})\\[2mm]
\hbox{ for }y_{i}\in Y_{i}, s_{1}\in I_{1}^{+}, z_{2}\in I_{2}+ \i I_{2}^{-}.
\end{array}
\]
In particular we obtain:
\[
\begin{array}{l}
\bar{F}_{2}( \i s_{2}, y_{2}, \is(s_{1}, y_{1}))=F(\i s_{1}, \i s_{2}, y_{1}, y_{2})\\[2mm]
\hbox{ for }y_{i}\in Y_{i}, \ s_{1}\in I_{1}^{+}, \ s_{2}\in I_{2}^{-}.
\end{array}
\]
By \eqref{e0.7d} this  gives:
\[
\begin{array}{l}
\big(K^{-1}\is\big)(s_{1}, y_{1}, -s_{2}, y_{2})= F(\i s_{1}, \i s_{2}, y_{1}, y_{2})\\[2mm]
\hbox{ for }y_{i}\in Y_{i}, \ s_{1}\in I_{1}^{+}, \ s_{2}\in I_{2}^{-},
\end{array}
\]
hence
\[
\begin{array}{l}
K^{-1}(s_{1}, y_{1}, s_{2}, y_{2})= F(\i s_{1}, \i s_{2}, y_{1}, y_{2})\\[2mm]
\hbox{ for }y_{i}\in Y_{i}, \ s_{1}\in I_{1}^{+}, \ s_{2}\in I_{2}^{-}.
\end{array}
\]
Using the fact that  $y^0_1,y^0_2$ were arbitrary we obtain the desired statement. \qed

\section{Stationary case}\label{sec:kms}

\subsection{KMS condition} We will now show some extra conclusions from our analysis in the special case when $K$ has coefficients independent of $s$. We work in the geometric setup summarized in 1)--7)  in Subsect.~\ref{ss:geometric}, with $\Me=\open{-\beta/2,\beta/2}\times\Sigma$. 


\begin{proposition}\label{propkms} Let $\beta>0$. Assume $\Sigma$ is real-analytic and $\rg$ is analytic, with coefficients independent of $t$, and assume $\mu$ is constant.  Let $\Me=\open{-\beta/2,\beta/2}\times\Sigma$, and let $\Lambda^\pm$ be defined as in Prop.~\ref{prop2pt} for some inverse $K^{-1}$ such that 
\beq   
\label{eq:per}
\lim_{s\to \beta/2-}(K^{-1}v)(s)=\lim_{s\to -\beta/2+}(K^{-1}v)(s), \ \ v\in H_{\cs}^{-2}(X).
\eeq
Then there exists $F_1\in\mo_{\rm temp}(\rr+\i 0;\cD'(\rr\times\Sigma^2))$ such that:
\beq\label{eq:simplecon1}
\bea
F_1(\i 0^+,y,t_2,y_2)&=\Lambda^+(0,y,t_2,y_2),\\
F_1(\i\beta^-,y,t_2,y_2)&=\Lambda^-(0,y,t_2,y_2),
\eea
\eeq
and $F_2\in\mo_{\rm temp}(\rr-\i 0;\cD'(\rr\times\Sigma^2))$ such that:
\beq\label{eq:simplecon2}
\bea
F_2(t_1,y_1,\i 0^-,y)&=\Lambda^-(t_1,y_1,0,y),\\
F_2(t_1,y_1,-\i\beta^+,y)&=\Lambda^+(t_1,y_1,0,y).
\eea
\eeq 
\end{proposition} 
\proof By Lemma \ref{lem:another} and its direct analogue with minuses replaced with pluses, there exist $F^\pm\in\mo_{\rm temp}(\rr\pm\i 0;\cD'(\rr\times\Sigma^2))$ such that for small $|t_1|$ and $|s_1|$,
\beq\label{eq:k1}
F^+(t_1+\i 0,y_1,t_2,y_2)=\Lambda^+(t_1,y_1,t_2,y_2),
\eeq
\beq\label{eq:k2}
F^+(\i s_1,y_1,t_2,y_2)=\big(K^{-1}\is\gamma^*U^*\big)(s_1,y_1,t_2,y_2), \ \ s_1>0,
\eeq
and
\beq\label{eq:k3}
F^-(t_1-\i 0,y_1,t_2,y_2)=\Lambda^-(t_1,y_1,t_2,y_2),
\eeq
\beq\label{eq:k4}
F^-(\i s_1,y_1,t_2,y_2)=\big(K^{-1}\is\gamma^*U^*\big)(s_1,y_1,t_2,y_2), \ \ s_1<0.
\eeq
Let us consider $K$ as a differential operator on $\rr\times\Sigma$ (rather than just on $X$). By \eqref{eq:per} we can extend the right hand side of \eqref{eq:k2} to a $\beta$-periodic distribution $w$ that solves $Kw=0$ in $\{ 0 <s_1  < \beta\}$. Let 
\[
w_{s_1}(y_1,t_2,y_2)\defeq \big( w(s_1,y_1,t_2,y_2),- \p_{s_1} w(s_1,y_1,t_2,y_2) \big)
\]
 and define
\[
\widetilde{F}^+(t_1+\i s_1, y_1, t_2, y_2)\defeq \big(U w_{s_1}\big)(t_1,y_1,t_2,y_2),
\]
where $U$ (see Subsect.~\ref{ss:tpf}) acts in the $y_1$ variables only. For the sake of simplicity, let us assume that $\rg=-dt^2+h$, with $h$ a (real) Riemmanian metric on $\Sigma$. We have
\beq\label{eq:wv1}
(\p_{t_1}^2-\Delta_{h(y_1)}+\mu)\widetilde{F}^+=0.
\eeq
Furthermore, $(-\p_{s_1}^2-\Delta_{h(y_1)}+\mu)\widetilde{F}^+$ also solves a wave equation in $(t_1,y_1)$, and it has vanishing Cauchy data if $s_1\in \open{0,\beta}$, hence
\beq\label{eq:wv2}
(-\p_{s_1}^2-\Delta_{h(y_1)}+\mu)\widetilde{F}^+=0
\eeq
for $s_1\in\open{0,\beta}$. Combining \eqref{eq:wv1} with \eqref{eq:wv2} we get $(\p_{t_1}^2+\p_{s_1}^2)\widetilde{F}^+=0$. This implies
\[
(\p_{t_1}^2+\p_{s_1}^2)\widetilde{F}^+_{\varphi}=0
\]
for
\[
\widetilde{F}^+_{\varphi}(s_1,t_1) \defeq \langle \widetilde{F}^+(s_1,t_1,\cdot), \varphi(\cdot)\rangle, \ \ \varphi\in\ccf(\rr\times\Sigma^2). 
\]
By ellipticity of $\p_{t_1}^2+\p_{s_1}^2$ we have $\widetilde{F}^+\in\mo(\rr+\i \open{0,\beta};\cD'(\rr\times\Sigma^2))$. This gives the analytic continuation of $F^+$. This argument can be easily extended to general $\rg$.


 We will prove that \eqref{eq:simplecon1} holds true with $F_1=\widetilde{F}^+$. The first identity in  \eqref{eq:simplecon1} is simply \eqref{eq:k1} with $y_1=y$, in the limit $t_1\to 0$ (which exists because $\Lambda^+$ solve the Klein-Gordon equation in $(t_1,y_1)$. Let us now prove the second identity in \eqref{eq:simplecon1}. Using \eqref{eq:k2}, then $\beta$-periodicity, and then \eqref{eq:k4} and \eqref{eq:k3}, we obtain
\[
\bea
F_1(\i \beta^-,y_1,t_2,y_2) &= w(\beta^-,y_1,t_2,y_2)\\
 &= \big(K^{-1}\is\gamma^*U^*\big)(0^-,y_1,t_2,y_2)\\
 &= F^-(\i 0^-,y_1,t_2,y_2)\\
 &= \Lambda^-(0,y_1,t_2,y_2).
\eea
\]   
The two identities \eqref{eq:simplecon2} are proved similarly.
\qed

The identities \eqref{eq:simplecon1}--\eqref{eq:simplecon2} are equivalent to the \emph{KMS condition} (recalled in Remark \ref{kmsremark} below) with respect to translations in $t$ for the quasi-free state associated with $\Lambda^\pm$.

\begin{corollary} Under the hypotheses of Prop.~\ref{propkms}, the state associated to the two-point functions $\Lambda^\pm$ is a $\beta$-KMS state. \end{corollary}

 We remark that this result can also be obtained (even without imposing ana\-ly\-ti\-city of the space-time) by more explicit, though more lengthy computations on the level of Cauchy data, see \cite{HHI,HHI2}.

\begin{remark}\label{kmsremark} Note that in our framework we have a pair of two-point functions $\Lambda^\pm$ and thus a pair of identities for the KMS condition: this is because we are using the terminology for charged fields. Recall that the abstract formulation of the KMS condition with respect to a given one-parameter group of automorphisms $\{ \tau_t \}$ is
\beq\label{eq:kms}
\omega\left(A \tau_{\i \beta}(B)\right)=\omega(BA),
\eeq
for all $A,B$, see e.g.~\cite[5.3.1]{BR} for the details. Here the (gauge-invariant) quasi-free state $\omega$ is defined from $\Lambda^\pm$ in such way that
\[
\langle \overline{v}, \Lambda^+ u \rangle = \omega\big(\psi(v)\psi^*(u)\big), \ \ \langle \overline{v}, \Lambda^- u \rangle =\omega\big(\psi^*(u)\psi(v)\big), \ \ v,u\in \ccf(M),
\]
in terms of the abstract charged fields $v\mapsto \psi^*(v),\psi(v)$, see \cite{analytic}. For such states it is sufficient to check \eqref{eq:kms} for all $A$ of the form $\psi(v)$ and $B$ of the form $\psi^*(u)$, and then for $A=\psi^*(u)$ and $B=\psi(v)$. Applied to $\tau_t\psi^*(v)=\psi^*(v(\cdot + t))$, this corresponds precisely to \eqref{eq:simplecon1} and \eqref{eq:simplecon2}. The more standard `neutral' terminology is recovered by re-writing everything in terms of the real fields $\phi(v)\defeq \frac{1}{\sqrt{2}}(\psi(v)+ \psi^{*}(v))$.
\end{remark}

\end{document}